\documentclass[lettersize,journal]{IEEEtran}
\usepackage{amsmath,amsfonts}
\usepackage{algorithmic}
\usepackage{algorithm}
\usepackage{array}
\usepackage[caption=false,font=normalsize,labelfont=sf,textfont=sf]{subfig}
\usepackage{textcomp}
\usepackage{stfloats}
\usepackage{url}
\usepackage{verbatim}
\usepackage{graphicx}
\usepackage{cite}
\usepackage{bm}
\usepackage{booktabs} 
\usepackage{xcolor}   
\usepackage{multirow} 
\begin{document}

\title{Low-Complexity MIMO Channel Estimation \\ with Latent Diffusion Models}

\author{Xiaotian Fan,~\IEEEmembership{Graduate Student Member,~IEEE}, Xingyu Zhou,~\IEEEmembership{Graduate Student Member,~IEEE}, \\Le Liang,~\IEEEmembership{Member,~IEEE}, and Shi Jin,~\IEEEmembership{Fellow,~IEEE}
\thanks{X. Fan, X. Zhou, L. Liang and S. Jin are with the School of Information Science and Engineering, Southeast University, Nanjing 210096, China (e-mail: xt\_fan@seu.edu.cn; xy\_zhou@seu.edu.cn; lliang@seu.edu.cn; jinshi@seu.edu.cn). L. Liang is also with Purple Mountain Laboratories, Nanjing 211111, China.}}

\maketitle

\begin{abstract}
Deep generative models offer a powerful alternative to conventional channel estimation by learning the complex prior distribution of wireless channels. Capitalizing on this potential, this paper proposes a novel channel estimation algorithm based on latent diffusion models (LDMs), termed posterior sampling with latent diffusion for channel estimation (PSLD-CE). The core of our approach is a lightweight LDM architecture specifically designed for channel estimation, which serves as a powerful generative prior to capture the intricate channel distribution. Furthermore, we enhance the diffusion posterior sampling process by introducing an effective approximation for the likelihood term and a tailored self-consistency constraint on the variational autoencoder latent space. Extensive experimental results demonstrate that PSLD-CE consistently outperforms a wide range of existing methods. Notably, these significant performance gains are achieved while maintaining low computational complexity and fast inference speed, establishing our method as a highly promising and practical solution for next-generation wireless systems.
\end{abstract}

\begin{IEEEkeywords}
Channel estimation, LDMs, posterior sampling, variational autoencoder, massive MIMO.
\end{IEEEkeywords}

\section{Introduction}
\IEEEPARstart{M}{assive} multiple-input multiple-output (MIMO) systems are fundamental to modern wireless communications, offering significant improvements in data throughput\cite{OVERMIMO}. However, achieving these performance gains relies heavily on precise channel state information (CSI). As antenna arrays grow in size, obtaining accurate CSI requires an increasingly large amount of pilot overhead. Consequently, the development of efficient channel estimation techniques that can provide accurate CSI with limited pilot resources is crucial for the practical deployment of massive MIMO.

Conventional channel estimation methods like least squares (LS) ignore prior channel information, resulting in poor performance with limited pilots. While linear minimum mean square error (LMMSE) estimation can significantly outperform LS by leveraging prior channel knowledge, it requires accurate channel covariance matrices that demand extensive pilot overhead to estimate and introduce substantial computational complexity.
To mitigate pilot overhead, numerous methods have been proposed to exploit the inherent sparsity of wireless channels. Many compressed sensing (CS) methods have been successfully applied to channel estimation, demonstrating excellent performance \cite{CS}. However, the simple assumption of sparsity often fails to capture the complex characteristics of real-world channels. Consequently, the performance of these CS-based methods can degrade dramatically in practical deployments.

As a representative technology in the new era of artificial intelligence, deep generative models have emerged as a powerful paradigm for learning complex data distributions. These capabilities make them particularly suitable for modeling complex wireless channel characteristics. For instance, the authors of \cite{VAE} proposed a channel estimation method that combines a variational autoencoder (VAE) with an LMMSE filter. The work in \cite{GAN} utilized generative adversarial networks (GANs) to capture the channel's prior distribution. Nevertheless, GANs are unstable to train, often suffering from mode collapse and requiring meticulous hyperparameter tuning.

Recently, diffusion models (DMs) have emerged as leading generative models, achieving state-of-the-art performance across diverse applications \cite{DDPM}. These models operate by progressively adding noise to data and then learning to reverse the process to generate samples following the data distribution. In \cite{HGM}, a DM-based posterior inference method was proposed for recovering channel information from low-resolution quantized measurements. In \cite{SGM}, score-based generative models (SGMs) were leveraged to perform posterior sampling for channel estimation. However, this approach utilized a network architecture originally designed for image processing, resulting in high model complexity and an excessively long sampling process, which does not meet the low-latency requirements of practical systems. In another work \cite{LC}, a low-complexity MIMO channel estimator was proposed based on DMs. However, its applicability is limited to scenarios with full and orthogonal pilots, which hampers its scalability. To alleviate the computational burden of DMs, latent diffusion models (LDMs) have been proposed \cite{LDM}. By performing the diffusion process in a lower-dimensional latent space, LDMs significantly reduce model complexity and accelerate the sampling process. 

This letter proposes using LDMs for high-dimensional channel estimation. We develop a lightweight and efficient LDM framework that employs a VAE to map channels into a latent space, where a simplified convolutional neural network (CNN) executes the diffusion process guided by a meticulously designed posterior inference scheme. Compared with existing diffusion model approaches, our design greatly reduces model parameters and floating-point operations (FLOPs) while preserving strong estimation accuracy. It performs particularly well under limited pilot resources and low SNR, making it suitable for large-scale MIMO systems.

\section{System Model and Preliminaries}

\subsection{MIMO Channel Estimation}
Consider a massive MIMO system where transmitter and receiver are equipped with $N_t$ and $N_r$ antennas, respectively. During the training phase, the transmitter sends pilot sequences of length $N_p$ to the receiver. As a result, the received signal $\mathbf{Y}\in\mathbb{C}^{N_p\times N_r}$ can be expressed as follows:
\begin{equation}\label{eq4.1}\mathbf{Y}=\mathbf{X}\mathbf{H}+\mathbf{N},\end{equation}
where \( \mathbf{H} \in \mathbb{C}^{N_t \times N_r} \) is the channel matrix, \( \mathbf{X} \in \mathbb{C}^{N_p \times N_t} \) denotes the pilot matrix, and \( \mathbf{N} \in \mathbb{C}^{N_p \times N_r} \) represents the noise matrix, with elements modeled as i.i.d. complex Gaussian random variables with zero mean and variance \( \sigma^2 \). 
\subsection{Latent Diffusion Models}
In the following, we briefly review the formulation of LDMs to provide the necessary foundation for the development of our proposed algorithm.

In LDMs, the training procedure follows a two-stage process. First, a VAE is trained to learn an efficient, low-dimensional representation of the data. Second, a diffusion model is trained in this latent space, with the VAE weights kept frozen during the process. For a given data sample $\mathbf{H}_0 \in \mathbb{R}^{C \times N_h \times N_w}$ from the input data distribution, the encoder of the VAE, denoted as $\mathcal{E}$, maps the input to a lower-dimensional latent representation $\mathbf{Z}_0 = \mathcal{E}(\mathbf{H}_0)$, where $\mathbf{Z}_0 \in \mathbb{R}^{C' \times N'_h \times N'_w}$. A key aspect of this stage is the spatial downsampling of the input by a factor $f = N_h / N'_h = N_w / N'_w$. In this work, we explore the impact of various downsampling factors, with detailed results presented in Section IV. Subsequently, the latent representation $\mathbf{Z}_0$ undergoes a forward diffusion process over $T$ timesteps, where Gaussian noise is progressively added according to a fixed variance schedule $\{\alpha_t\}_{t=1}^T$:
\begin{equation}
\begin{aligned}
    \mathbf{Z}_t &= \sqrt{\alpha_t}\mathbf{Z}_{t-1} + \sqrt{1 - \alpha_t}\bm{\epsilon}_{t-1} \\
                &= \sqrt{\bar{\alpha}_t}\mathbf{Z}_0 + \sqrt{1 - \bar{\alpha}_t}\bm{\epsilon}_0,
\end{aligned}
\quad \text{for } t=1, \dots, T,
\label{eq:forward_process}
\end{equation}
where $\bm{\epsilon}_{t-1} \sim \mathcal{N}(\mathbf{0}, \mathbf{I})$ denotes a standard Gaussian random matrix, and $\bar{\alpha}_t = \prod_{i=1}^t \alpha_i$.

The core of LDMs is to learn the reverse process, $p_{\bm{\theta}}(\mathbf{Z}_{t-1}|\mathbf{Z}_t)$, to approximate the true but intractable reverse distribution, $p(\mathbf{Z}_{t-1}|\mathbf{Z}_t)$. The key insight is that this distribution becomes a tractable Gaussian when conditioned on the initial data $\mathbf{Z}_0$:
\begin{equation}
    p(\mathbf{Z}_{t-1}|\mathbf{Z}_t, \mathbf{Z}_0) = \mathcal{N}(\mathbf{Z}_{t-1}; \tilde{\bm{\mu}}_t(\mathbf{Z}_t, \mathbf{Z}_0), \tilde{\beta}_t\mathbf{I}),
\end{equation}
where $\tilde{\bm{\mu}}_t$ is the mean and $\tilde{\beta}_t$ is the variance given by $\tilde{\beta}_t = \frac{1 - \bar{\alpha}_{t-1}}{1 - \bar{\alpha}_t} \beta_t$, with \( \beta_t = 1 - \alpha_t \). Since the target distribution is Gaussian, it is natural to parameterize $p_{\bm{\theta}}$ as a Gaussian with a learnable mean:
$
    p_{\bm{\theta}}(\mathbf{Z}_{t-1}|\mathbf{Z}_t) = \mathcal{N}(\mathbf{Z}_{t-1}; \bm{\mu}_{\bm{\theta}}(\mathbf{Z}_t, t), \sigma_t^2\mathbf{I}).
$
The objective is thus to train the network $\bm{\mu}_{\bm{\theta}}$ to predict $\tilde{\bm{\mu}}_t$.

Following \cite{DDPM}, instead of directly training the mean $\bm{\mu}_{\bm{\theta}}$, a more effective approach is to further parameterize the model to predict the noise component $\bm{\epsilon}_t$ from the forward process equation:
$
\mathbf{Z}_t = 
\sqrt{\bar{\alpha}_t}\mathbf{Z}_0 + 
\sqrt{1-\bar{\alpha}_t}\bm{\epsilon}_t.
$
A denoising network $\bm{\epsilon}_{\bm{\theta}}(\mathbf{Z}_t, t)$ is trained to estimate this noise. This strategy is effective because the true mean $\tilde{\bm{\mu}}_t$ can be expressed as a function of $\mathbf{Z}_t$ and $\bm{\epsilon}$. 

Once trained, new samples are generated by iteratively applying the reverse step from \( t=T \) down to 1:
\begin{equation}
    \mathbf{Z}_{t-1} = \frac{1}{\sqrt{\alpha_t}} \left( \mathbf{Z}_t - \frac{1-\alpha_t}{\sqrt{1-\bar{\alpha}_t}} \bm{\epsilon}_{\bm{\theta}}(\mathbf{Z}_t, t) \right) + \sigma_t \mathbf{n},
\end{equation}
where \( \mathbf{n} \) is a standard Gaussian noise, and \( \sigma_t^2 \) is often set to \( \tilde{\beta}_t \). Finally, the generated latent sample $\mathbf{Z}_0$ is passed through the VAE's decoder, $\mathcal{D}$, to reconstruct the final data sample $\hat{\mathbf{H}} = \mathcal{D}(\mathbf{Z}_0)$, which follows the original data distribution.
\begin{figure*}[!t]
\centering
\includegraphics[width=5.6in]{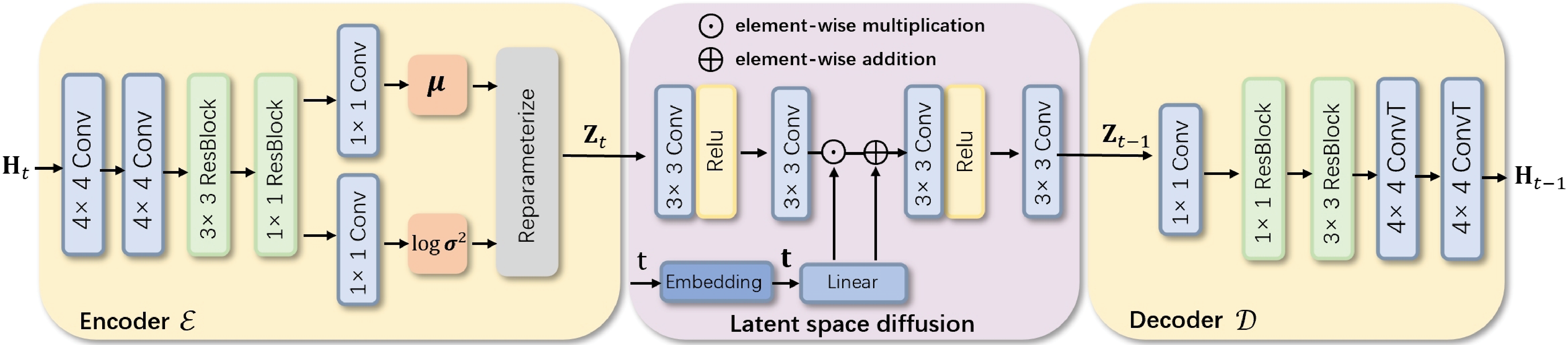}
\caption{LDM architecture with a lightweight VAE and a time-embedded CNN for the DM.}
\label{fig_0}
\end{figure*}

\section{Proposed Methods}
Following the training scheme of LDMs, this section is divided into two parts. We begin by designing a VAE for compressing high-dimensional channel data, and subsequently, we integrate this pre-trained VAE to develop the LDM-based channel estimation framework.
\subsection{VAE Architecture for Latent Channel Compression}
In the VAE architecture, an encoder $\mathcal{E}$ first maps the high-dimensional channel data $\mathbf{H}$ into a structured, lower-dimensional latent representation $\mathbf{Z}$. Subsequently, a decoder $\mathcal{D}$ reconstructs the original data from this latent representation.

\subsubsection{Network Architecture}
As illustrated in Fig.~\ref{fig_0}, the encoder $\mathcal{E}$ initiates by downsampling the spatial dimensions of the input channel using two convolutional layers. These are followed by two residual blocks. Each block consists of a convolutional layer with skip connection, followed by a ReLU activation function. This structure serves to refine the extracted features without increasing network depth. Finally, two parallel $1 \times 1$ convolutional layers act as output heads to parameterize the variational posterior distribution $q(\mathbf{Z}|\mathbf{H})$ by producing its mean $\bm{\mu}$ and log-variance $\log\bm{\sigma}^2$ \cite{VAE2}. The final latent variable $\mathbf{Z}$ is then obtained by applying the reparameterization trick to these two outputs. The decoder $\mathcal{D}$ symmetrically mirrors the encoder's architecture, employing transposed convolutional layers to progressively upsample the features and reconstruct the original input $\mathbf{H}$ from the latent representation $\mathbf{Z}$.

\subsubsection{Key Design Considerations}
The training objective of the VAE is designed to address the challenge of an intractable posterior \( p(\mathbf{Z}|\mathbf{H}) \). 
To overcome this, the VAE introduces an approximate posterior \( q(\mathbf{Z}|\mathbf{H}) \), parameterized by the encoder network, and aims to minimize the Kullback-Leibler (KL) divergence between these two distributions. 
This is achieved by maximizing the Evidence Lower Bound (ELBO) \cite{VAE2}. A weight \( \lambda \) is commonly introduced, resulting in a modified ELBO defined as:
\begin{equation}
    \mathcal{L}_{\text{VAE}} = \mathbb{E}_{q(\mathbf{Z}|\mathbf{H})}[\log p(\mathbf{H}|\mathbf{Z})] - \lambda \cdot D_{\text{KL}}(q(\mathbf{Z}|\mathbf{H}) \,||\, p(\mathbf{Z})).
    \label{eq:vae_loss}
\end{equation}
The objective function consists of two key components. 
The first term is the reconstruction term, which encourages the decoder \( p(\mathbf{H}|\mathbf{Z}) \) to accurately reconstruct the input data. 
The second term is a regularizer that minimizes the KL divergence between the approximate posterior and the prior distribution \( p(\mathbf{Z}) \), typically a standard Gaussian distribution. 

In commonly accepted practices, the weight of the KL-divergence term \( \lambda \) is set to 1, at which point $\mathcal{L}_{\text{VAE}}$ becomes the standard ELBO. This encourages the latent space to adopt a smooth, Gaussian-like structure, which is beneficial for generative tasks. However, this strong regularization often comes at the cost of reconstruction fidelity, leading to overly smooth or blurry outputs. 
In contrast, within the LDM framework, the VAE acts as a high-fidelity autoencoder rather than a generator, as the generative task is delegated to the diffusion model. 
To this end, we weaken the regularization by setting the KL divergence weight to a small value ($\lambda = 10^{-7}$). 
While minimal, this non-zero weight imposes a subtle regularization, which encourages a smoother and more structured latent manifold.
This strategic choice ensures high fidelity, providing a smooth and information-rich latent representation that is crucial for the subsequent diffusion process.

\subsection{Latent Space Diffusion for Channel Estimation}
In the preceding subsection, we developed a VAE to compress the high-dimensional channel $\mathbf{H}$ into a low-dimensional latent variable $\mathbf{Z} = \mathcal{E}(\mathbf{H})$. All subsequent diffusion processing is then performed in this latent space.

Our objective is to frame channel estimation as a Bayesian inference problem, where we aim to sample from the posterior distribution $p(\mathbf{Z}|\mathbf{Y})$. We achieve this by guiding the reverse diffusion process with the received signal $\mathbf{Y}$, which yields the update rule for each sampling step \cite{DPS}:
\begin{equation}
\label{6}
\mathbf{Z}_{t-1} = \frac{1}{\sqrt{\alpha_t}} \left( \mathbf{Z}_t + (1 - \alpha_t) \nabla_{\mathbf{Z}_t} \log p(\mathbf{Z}_t|\mathbf{Y}) \right),
\end{equation}
where $\nabla_{\mathbf{Z}_t} \log p(\mathbf{Z}_t|\mathbf{Y})$ can be decomposed using Bayes' rule:
\begin{equation}
\nabla_{\mathbf{Z}_t} \log p(\mathbf{Z}_t|\mathbf{Y}) = \nabla_{\mathbf{Z}_t} \log p(\mathbf{Z}_t) + \nabla_{\mathbf{Z}_t} \log p(\mathbf{Y}|\mathbf{Z}_t).
\end{equation}
Next, we explain how to estimate these two terms and construct the complete posterior sampling algorithm.

\subsubsection{Prior score} We leverage the pre-trained LDM as a powerful prior for the channel distribution. The corresponding prior score is derived directly from the model's noise prediction network:
\begin{equation}
\nabla_{\mathbf{Z}_t} \log p(\mathbf{Z}_t) \approx - \frac{1}{\sqrt{1 - \bar{\alpha}_t}} \bm{\epsilon}_{\bm{\theta}}(\mathbf{Z}_t, t).
\end{equation}
This term guides the sampling towards the learned manifold of valid channel representations.

\subsubsection{Likelihood score} The likelihood score steers the sampling process towards solutions consistent with the received signal $\mathbf{Y}$. The core challenge is that the physical model defines a relationship between $\mathbf{Y}$ and the clean latent variable $\mathbf{Z}_0$, not the noisy intermediate $\mathbf{Z}_t$. The exact likelihood is therefore an intractable integral over all possible clean latents:
$
p(\mathbf{Y}|\mathbf{Z}_t) = \mathbb{E}_{\mathbf{Z}_0 \sim p(\mathbf{Z}_0|\mathbf{Z}_t)}[p(\mathbf{Y}|\mathbf{Z}_0)].
$
To overcome this, we introduce a key approximation. Instead of directly integrating over $p(\mathbf{Z}_0|\mathbf{Z}_t)$, we approximate it using its expectation, denoted as $\hat{\mathbf{Z}}_0 := \mathbb{E}_{\mathbf{Z}_0 \sim p(\mathbf{Z}_0 | \mathbf{Z}_t)} [\mathbf{Z}_0]$. This leads to the tractable approximation:
$
p(\mathbf{Y}|\mathbf{Z}_t) \approx p(\mathbf{Y}|\hat{\mathbf{Z}}_0),
$
where $\hat{\mathbf{Z}}_0$ represents the denoised estimate of $\mathbf{Z}_0$ given $\mathbf{Z}_t$ \cite{DPS}. It can be directly calculated using the noise prediction network:
\begin{equation}
\hat{\mathbf{Z}}_0 = \frac{1}{\sqrt{\bar{\alpha}_t}} (\mathbf{Z}_t - \sqrt{1 - \bar{\alpha}_t} \bm{\epsilon}_{\bm{\theta}}(\mathbf{Z}_t, t)).
\end{equation}
Then we map this estimated latent variable back to the channel space, $\hat{\mathbf{H}}_0 = \mathcal{D}(\hat{\mathbf{Z}}_0)$, and compute the gradient of the log-likelihood. Under a Gaussian noise assumption, this results in the gradient of the reconstruction error \cite{DPS}:
\begin{equation}
\nabla_{\mathbf{Z}_t} \log p(\mathbf{Y}|\mathbf{Z}_t) \approx -\nabla_{\mathbf{Z}_t} \left( \| \mathbf{Y} - \mathbf{X} \mathcal{D}(\hat{\mathbf{Z}}_0) \|^2_2 \right).
\end{equation}
However, relying solely on this approximation is insufficient for a LDM. The VAE encoder performs a many-to-one mapping, where multiple latent variables $\mathbf{Z}$ may correspond to the same observation $\mathbf{Y}$. As a result, the likelihood gradient alone may cause conflicting updates and instability.

To resolve this ambiguity, we introduce a self-consistency regularization term \cite{PSLD}. The key idea is to enforce that the solution lies on the manifold of valid latent variables learned by the VAE. This is achieved by penalizing estimates $\hat{\mathbf{Z}}_0$ that are not fixed points of the decoder-encoder composition (i.e., where $\hat{\mathbf{Z}}_0 \neq \mathcal{E}(\mathcal{D}(\hat{\mathbf{Z}}_0))$). The final, stabilized likelihood score is composed of two terms:
\begin{equation}\begin{aligned}
&\nabla_{\mathbf{Z}_t}\log p(\mathbf{Y}|\mathbf{Z}_t)  \approx -\eta\nabla_{\mathbf{Z}_t} \left( \| \mathbf{Y} - \mathbf{X} \mathcal{D}(\hat{\mathbf{Z}}_0) \|^2_2 \right) \\
 & -\gamma\nabla_{Z_t}\left|\left|\hat{\mathbf{Z}}_0-\mathcal{E}\Bigl(\mathbf{Y}\mathbf{X}^T+\mathcal{D}(\hat{\mathbf{Z}}_0)(\boldsymbol{I}-\mathbf{X}\mathbf{X}^T)\Bigr)\right|\right|_2^2,
\end{aligned}\end{equation}
where the first term ensures consistency with $\mathbf{Y}$, while the second self-consistency term ensures the latent variable is a valid representation. $\eta$ and $\gamma$ are scaling factors to balance the two terms, where $\gamma$ is typically set to one-tenth of $\eta$.

The complete posterior sampling is performed through an iterative reverse diffusion process. At each step, the update of Eq.~(\ref{6}) is guided by the combination of the prior score and the stabilized likelihood score. This entire procedure, which we term posterior sampling with latent diffusion for channel estimation (PSLD-CE), effectively samples from the posterior distribution and is summarized in Algorithm~\ref{alg:psld}.

\begin{algorithm}[t]
    \caption{PSLD-CE}
    \label{alg:psld}
    \begin{algorithmic}[1] 
        \STATE \textbf{Input:} $\mathbf{Y}$, $\mathbf{X}$, $\eta$, $\gamma$, $\mathcal{E}$, $\mathcal{D}$, $\bm{\epsilon}_{\bm{\theta}}$

        \STATE $\mathbf{Z}_T \sim \mathcal{N}(\mathbf{0}, \mathbf{I})$
        
        \FOR{$t = T$ \textbf{to} $1$} 
            \STATE $\hat{\bm{\epsilon}} \leftarrow \bm{\epsilon}_{\bm{\theta}}(\mathbf{Z}_t, t)$
            \STATE $\hat{\mathbf{Z}}_0 \leftarrow \frac{1}{\sqrt{\bar{\alpha}_t}} (\mathbf{Z}_t - \sqrt{1 - \bar{\alpha}_t} \hat{\bm{\epsilon}})$
            \STATE $\mathbf{Z}'_{t-1} \leftarrow  \frac{1}{\sqrt{\alpha_t}} \left( \mathbf{Z}_t - \frac{1-\alpha_t}{\sqrt{1-\bar{\alpha}_t}} \hat{\bm{\epsilon}} \right)$
            \STATE $\mathbf{Z}''_{t-1} \leftarrow \mathbf{Z}'_{t-1} - \eta \frac{(1 - \alpha_t)}{\sqrt{\alpha_t}}\nabla_{\mathbf{Z}_t} \|\mathbf{Y} - \mathbf{X}(\mathcal{D}(\hat{\mathbf{Z}}_0))\|_2^2$
            \STATE $\mathbf{Z}_{t-1} \leftarrow \mathbf{Z}''_{t-1} - \gamma \frac{(1 - \alpha_t)}{\sqrt{\alpha_t}} \nabla_{\mathbf{Z}_t} \|\hat{\mathbf{Z}}_0-\mathcal{E}\Bigl(\mathbf{Y}\mathbf{X}^T+\mathcal{D}(\hat{\mathbf{Z}}_0)(\boldsymbol{I}-\mathbf{X}\mathbf{X}^T)\Bigr)\|_2^2$
        \ENDFOR 
        
        \RETURN $\hat{\mathbf{H}} = \mathcal{D}(\mathbf{Z}_0)$ 
    \end{algorithmic}
\end{algorithm}

\section{Simulation Results}
The channel models employed in this study are based on the 3GPP clustered delay line (CDL) model, generated via the MATLAB 5G Toolbox. The number of antennas at the transmitter and receiver is set as $N_t$ = 64 and $N_r$ = 16, respectively.
We generated a dataset of 10,000 independent channel realizations for training, 1,000 for validation, and 1,000 for testing.

The performance metric for channel estimation is the normalized mean-squared error (NMSE), which is defined as $\mathbb{E}\big\{\big\Vert\mathbf{H}-\hat{\mathbf{H}}\big\Vert_2^2\big\} / \mathbb{E}\big\{\big\Vert\mathbf{H}\big\Vert_2^2\big\}$.

\subsection{Selecting the Latent Dimension}
\begin{figure}[!t]
\centering
\includegraphics[width=2.75in]{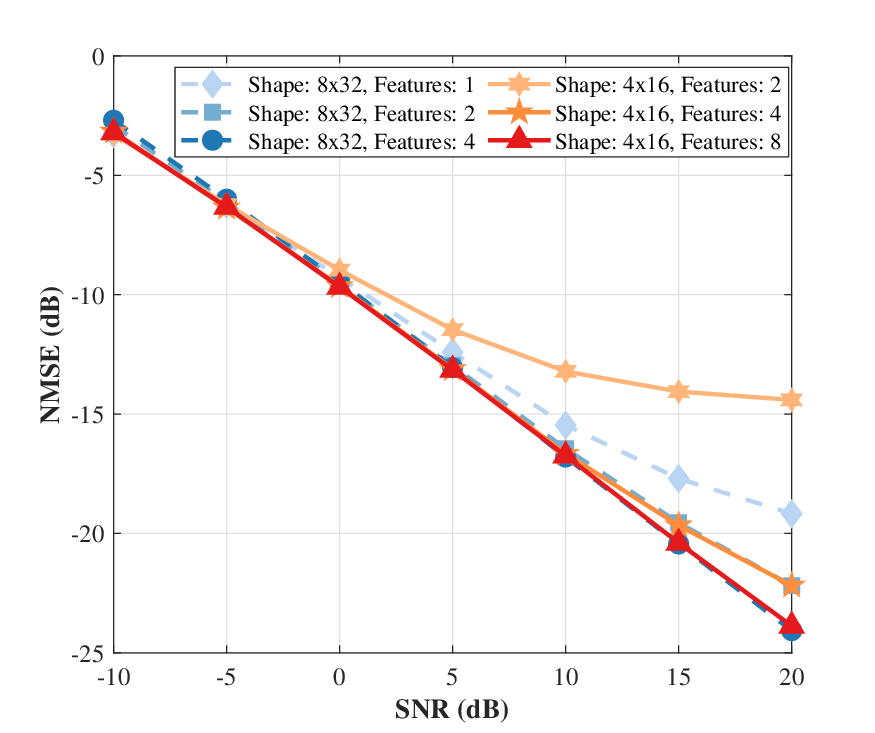}
\caption{NMSE of channel estimation with respect to different latent dimensions of VAE.}
\label{fig_1}
\end{figure}

The selection of the VAE's latent dimension is a critical step that achieves a fundamental trade-off: higher compression reduces computational complexity, while lower compression preserves the information essential for estimation accuracy. To identify an appropriate configuration, we evaluated both a VAE and a vector-quantized VAE (VQ-VAE) across various latent dimensions, with the recovery NMSE as the performance metric. As shown in Table~\ref{tab:model_comparison_optimized} and Fig.~\ref{fig_1}, our results identify the VAE with a latent dimension of \( 8 \times 4 \times 16 \) as the best configuration, where 8 corresponds to the number of features and $4 \times 16$ to the spatial dimensions of the latent space.
This latent dimension achieves the best performance in channel reconstruction fidelity, while also offering a compression ratio that significantly reduces the computational complexity. Consequently, this configuration is adopted in the remainder of this work.

\begin{table}[htbp]
  \centering
  \caption{Performance of VAE with various latent dimensions}
  \label{tab:model_comparison_optimized}
  \renewcommand{\arraystretch}{0.8} 
  \setlength{\tabcolsep}{7pt} 
  \begin{tabular}{cccc}
    \toprule 
    \textbf{Model} & \textbf{Latent Dim} & \textbf{Compression Ratio} & \textbf{NMSE (dB)} \\
    \midrule
    \multirow{6}{*}{VAE}    
      & $2\times4\times16$ & 16 & -15.15 \\
      & $4\times4\times16$ & 8 & -26.76 \\ 
      & $8\times4\times16$ & 4 & \textbf{-34.07} \\ 
      & $1\times8\times32$ & 8 & -21.07 \\
      & $2\times8\times32$ & 4 & -27.41 \\
      & $4\times8\times32$ & 2 & -33.36 \\ 
    \midrule 
    \multirow{2}{*}{VQ-VAE} 
      & $4\times4\times16$ & 8 & -18.70 \\
      & $8\times4\times16$ & 4 & -17.31 \\
    \bottomrule 
  \end{tabular}
\end{table}

\subsection{Channel Estimation}
We compare our proposed PSLD-CE with the following baseline methods: \textbf{LMMSE:} Linear estimator using the sample covariance computed via all training samples.
    \textbf{Orthogonal matching pursuit (OMP):} A greedy algorithm iteratively selects dictionary atoms most correlated with the signal to approximate the target signal. \textbf{Fast iterative shrinkage-thresholding algorithm (FISTA):} An accelerated optimization method, achieving faster convergence by combining iterative shrinkage-thresholding steps with a momentum-based acceleration technique. 
    \textbf{EM-GM-AMP:} The approximate message passing (AMP)-based iterative CS scheme from, which adopts a Gaussian mixture prior for angular domain channels.
    \textbf{LDAMP:} A supervised DL-based channel estimator by unfolding the AMP method and incorporating a denoising CNN \cite{LDAMP}.
    \textbf{VAE:} The VAE-based channel estimator that models the channel distribution as conditionally Gaussian \cite{VAE}. 
    \textbf{SGM:} The SGM-based estimator, which learns a score-based generative prior and performs estimation using annealed Langevin dynamics \cite{SGM}.
    \textbf{Low-complexity DM:} A lightweight MIMO channel estimator using the DM as a generative prior \cite{LC}.

\begin{figure}[!t]
\centering
\includegraphics[width=2.68in]{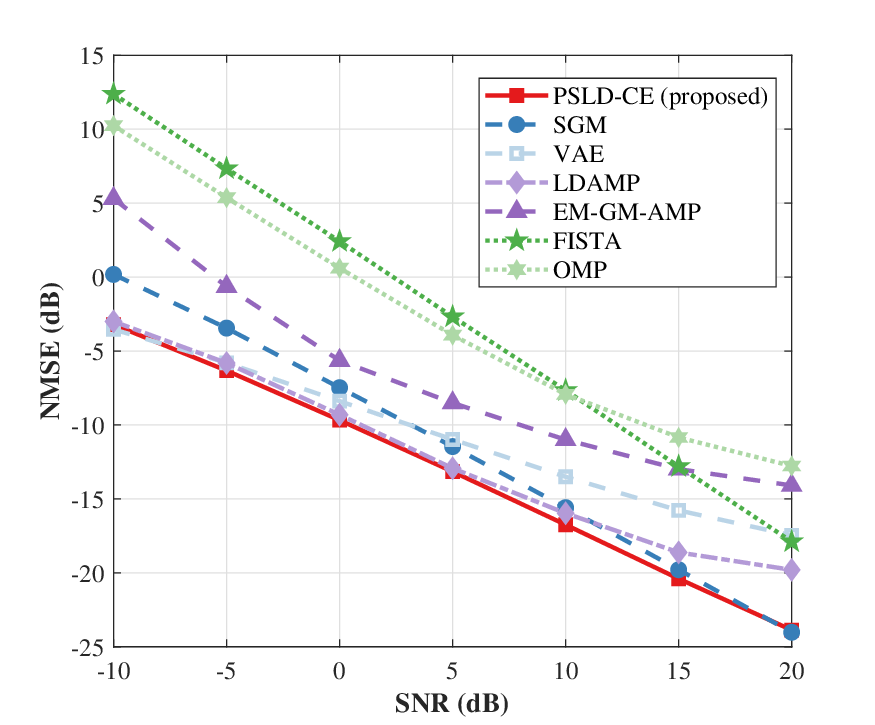}
\caption{The NMSE versus the SNR with QPSK pilot when $N_p/N_t = 0.6$.}
\label{fig_2}
\end{figure}

We first evaluate the performance of our proposed PSLD-CE method in an underdetermined scenario, employing $N_p = 38$ pilots whose symbols are randomly drawn from a quadrature phase-shift keying (QPSK) constellation.
As illustrated in Fig. \ref{fig_2}, the proposed method demonstrates a significant performance advantage over all compared schemes. The CS based algorithms, namely OMP and FISTA, exhibit suboptimal performance. This is attributed to the fact that the generated channel dataset does not strictly adhere to the sparsity assumption in the angular domain. By employing a more sophisticated Gaussian mixture model prior, the EM-GM-AMP algorithm achieves an improvement over conventional CS methods; however, a significant performance gap still exists when compared to our proposed approach. Furthermore, the deep learning-based method, LDAMP, achieves performance comparable to our proposed method at low signal-to-noise ratios (SNRs). However, it experiences a significant performance degradation in the high SNR regime, resulting in a performance gap of approximately 5\,dB targeting the NMSE of -20\,dB. Furthermore, the SGM algorithm is also outperformed by our method across the entire SNR range.

\begin{figure}[!t]
\centering
\includegraphics[width=2.9in]{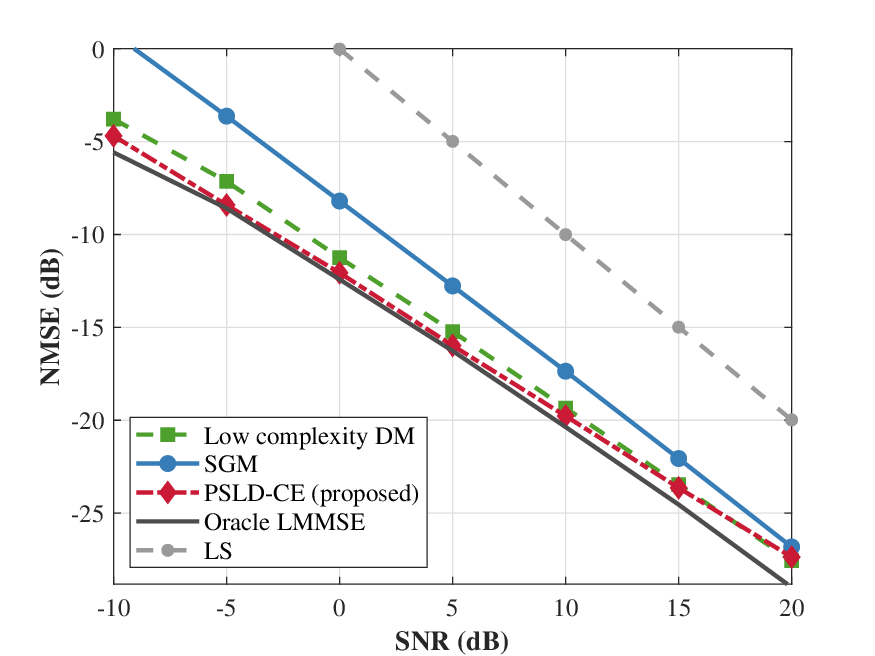}
\caption{The NMSE versus the SNR with DFT pilot when $N_p/N_t = 1$.}
\label{fig_3}
\end{figure}

Furthermore, our analysis considers a fully-determined scenario where $N_p = N_t$, and using a discrete Fourier transform (DFT) matrix as the unitary pilot matrix.
In Fig. \ref{fig_3}, we compare the performance of several diffusion-based methods against the classical LS and the LMMSE estimators. The results indicate that the conventional LS estimator performs poorly due to the absence of prior channel knowledge. In contrast, our proposed PSLD-CE method exhibits robust performance across the entire range of SNRs. Notably, its accuracy closely approaches that of the LMMSE method, a powerful benchmark that utilizes the channel covariance matrix estimated from the training dataset. Furthermore, our PSLD-CE method also maintains a substantial advantage over the SGM algorithm. Compared to the Low-Complexity DM algorithm, our proposed method demonstrates superior performance at low SNRs while requiring lower computational complexity, as verified in Table~\ref{tab:complexity_comparison}. Moreover, a key strength of our proposed algorithm lies in its robustness to challenging pilot conditions. It maintains high performance even with non-orthogonal and insufficient pilot sequences, as observed in Fig. \ref{fig_2}. 

\begin{table}[t]
\centering
\caption{Model Size and Inference FLOPs Comparison}
\label{tab:complexity_comparison} 
\renewcommand{\arraystretch}{0.8}
\begin{tabular}{@{}lcc@{}} 
\toprule
\textbf{Algorithm} & \multicolumn{1}{c}{\textbf{Model Size}} & \multicolumn{1}{c}{\textbf{FLOPs}} \\
                   & \multicolumn{1}{c}{\textbf{(Params.)}}   & \multicolumn{1}{c}{\textbf{(Inference)}} \\
\midrule
LDAMP \cite{LDAMP}      & 4.81 M &  0.699 G \\
VAE \cite{VAE}             & 8.95 M & 5.907 G \\ 
SGM \cite{SGM}          & 5.89 M & 2918 G \\
Low-Complexity DM \cite{LC} &  0.05 M & 3.103 G \\
\midrule
\textbf{PSLD-CE (Proposed)} & \textbf{0.2 M} & \textbf{1.615 G} \\ 
\bottomrule
\end{tabular}
\end{table}

Table~\ref{tab:complexity_comparison} compares the model size and inference FLOPs of our proposed PSLD-CE with several deep learning baselines. Our proposed PSLD-CE is highly compact, containing only 0.2~million parameters (0.16~M for the VAE and 0.04~M for the DM), rendering it over 24 times smaller than deep learning-based methods like SGM (5.89~M), VAE (8.95~M), and LDAMP (4.81~M). For inference, PSLD-CE requires approximately 1.615~GFLOPs. This efficiency is achieved by performing the DM update within a compressed latent space, which reduces the computational cost of the DM module to just 0.005~GFLOPs per step. In contrast, the Low-Complexity DM requires significantly higher computation cost. This method initializes with an LS estimation and then employs a variable number of reverse diffusion steps depending on the SNR. We average its computational cost across SNRs from -10~dB to 20~dB, resulting in 3.103 GFLOPs. This high cost arises primarily from operating directly on full-dimensional data, resulting in substantial per-step computations of 0.107 GFLOPs—over 20 times greater than our approach.

\section{Conclusion}
This work proposed a novel channel estimation framework, PSLD-CE, which effectively leverages the generative power of LDMs. We developed a lightweight LDM architecture and an enhanced posterior sampling technique, which incorporates a tailored likelihood approximation and a VAE-based regularization. Experimental results demonstrate that PSLD-CE achieves superior channel estimation accuracy compared to existing algorithms while maintaining exceptional efficiency in model size and inference speed, marking a significant step toward practical, high-performance channel estimation solutions.

\bibliographystyle{IEEEtran}

\end{document}